\documentclass{article}
\usepackage{amssymb}
\usepackage{amsthm}
\usepackage{amsmath}

\newtheorem{lemma}{Lemma}
\newtheorem{theorem}{Theorem}
\newtheorem{corollary}{Corollary}
\newtheorem{example}{Example}

\begin{document}
\title{Permutation Arrays Under the Chebyshev Distance \thanks{The research was supported in part by the
National Science Council of Taiwan under contracts  NSC-95-2221-E-009-094-MY3, NSC-96-2221-E-009-026,
NSC-96-3114-P-001-002-Y and NSC-96-2219-E-009-013 and by the Norwegian Research Council.}}
\author{
Torleiv Kl{\o}ve,\thanks{T. Kl{\o}ve is with the Department of Informatics, University of Bergen, N-5020 Bergen, Norway
(Email: Torleiv.Klove@ii.uib.no).}
Te-Tsung Lin, Shi-Chun Tsai, and  Wen-Guey~Tzeng
\thanks{T.-T. Lin, S.-C. Tsai and W.-G. Tzeng are with the Department of Computer Science,
 National Chiao Tung University,
 Hsinchu 30050, Taiwan
(Email: atman.cs94g@nctu.edu.tw, sctsai@csie.nctu.edu.tw, wgtzeng@cs.nctu.edu.tw).
}}

\maketitle

\begin{abstract}
An $(n,d)$ permutation array (PA) is a subset of $S_n$ with the property that the distance (under some
metric) between any two permutations in the array is at least $d$.  They became popular recently
for communication over power lines. Motivated by an application to flash memories, in this paper the metric used is the Chebyshev metric.
A number of different constructions are given as well as bounds on the size of such PA.
\end{abstract}

\section{Introduction}\label{introduction:sec}

Let $S_n$ denote the set of all permutations of length $n$. A permutation array of length $n$ is a subset of $S_n$.
Recently, Jiang et. al \cite{Jiang1,Jiang2} showed an interesting new application of permutation arrays for flash memories, where
they used different distance metrics to investigate efficient rewriting schemes.  Under the multi-level flash memory model,
we find the Chebyshev metric very appropriate for studying the recharging and error correcting issues. We note it by
$d_{\max}$. For $\pi, \sigma \in S_n$, $d_{\max}(\pi,\sigma)= \max_i |\pi_i - \sigma_i|$.
 We consider a noisy channel where pulse
amplitude modulation (PAM) is used with different amplitude levels for each permutation symbol.  The noise in the
channel is an independent Gaussian distribution with zero mean for each position. The received sequence is the
original permutation distorted by Gaussian noise, and its ranking can be seen as a permutation, which can be
different from the original one.

To study the correlations between ranks, several metrics
on permutations were introduced, such as the Hamming distance, the minimum number of transpositions taking one
permutation to another, etc. \cite{Diaconis}, \cite{Kendall90}. For instance, Stoll and Kurz \cite{SK68}
investigated a detection scheme of permutation arrays using Spearman's rank correlation.  Chadwick and Kurz
\cite{CK69} studied the permutation arrays based on Kendall's tau.

Under the model of additive white Gaussian noise (AWGN) \cite{Haykin}, there is
only a small probability for any amplitude level to deviate significantly from the original one. This inspired us
to use the Chebyshev distance.
Observe that two permutations with a large Hamming
distance can actually have a small Chebyshev distance and vice versa.  They appear to complement each
other in some sense.

In this paper, we give a number of constructions of  PAs. For some we give efficient decoding algorithms.
We also consider encoding from vectors into permutations.

\section{Notations}

We use $[n]$ to denote the set $\{1,\dots,n\}$. $S_n$ denotes the set of all
permutations of $[n]$. For any set $X$, $X^n$ denotes the set of all $n$-tuples with elements from $X$.

Let
$\iota$ denote the identity permutation in $S_n$. The Chebyshev
distance between two permutations $\pi,\sigma\in S_n$ is
\[d_{\max}(\pi,\sigma) = \max\{ |\pi_j - \sigma_j|\mid 1\le j \le n\}.\]

An $(n,d)$ permutation array (PA) is a subset of $S_n$ with the property that the Chebyshev distance between any two
distinct permutations in the array is at least $d$. We sometimes refer to the elements of a PA as code words.

The maximal size of an $(n,d)$ PA is denoted by $P(n,d)$.
Let $V(n,d)$ denote the number of permutations in $S_n$ within Chebyshev distance $d$ of the identity permutation.
Since $d_{\max}(\iota,\sigma)=d_{\max}(\pi,\pi\sigma)$, the number of permutations in $S_n$ within Chebyshev distance $d$ of
any permutation $\pi\in S_n$ will also be $V(n,d)$. Bounds on $P(n,a)$ and $V(n,d)$ will be considered in Sec. \ref{sec-bounds}.


\section{Constructions}
\label{sec-const}

In this section we give a number of constructions of PAs, one explicit and some recursive.

\subsection{An explicit construction}
Let $n$ and $d$ be given.  Define
\[C=\{(\pi_1,\dots, \pi_n)\in S_n| \pi_i\equiv i \pmod d \mbox{ for all } i \in [n]\}.\]
If $n=ad+b$, where $0\le b < d$, then $C$ is an $(n, d)$ PA and
\[|C|= ((a+1)!)^b(a!)^{d-b}. \]
In particular, we get the following bound.
\begin{theorem}\label{expl}
If $n=ad+b$, where $0\le b < d$, then
\[P(n,d)\ge ((a+1)!)^b(a!)^{d-b}.\]
\end{theorem}

\begin{example}
\label{expb}
For $d=2$, we get
\[P(2a,2)\ge (a!)^2.\]
\end{example}

We note that if $2d > n$, then $a=1$ and $b=n-d$ and so $|C|=2^{n-d}$.
If $2d=n$, then $a=2$, $b=0$, and we have $|C|=2^d=2^{n-d}$ as well.
However, if $2d <n$, then $|C|> 2^{n-d}$.
Especially, when $d$ is small relative to $n$, $|C|$ is much larger than $2^{n-d}$.  For example,
for $n=30, d=2$, $|C|/2^{n-d}\approx 6.37 \times 10^{15}.$

This construction has a very simple decoding algorithm.
For $d\ge 2t+1$, we can correct error up to size $t$ in any coordinate.
For coordinate $i$, the codeword has value $\pi_i\equiv i \pmod d$.
Suppose that this coordinate is changed into $\sigma=\pi_i+u$, where
$|u|\le t$.  Then $\pi_i$ is the integer congruent to $i$ which is closest
to $\sigma$.  Therefore, decoding of position $i$ is done by first computing
\[a\equiv i-\sigma  \pmod d,\]
 where $ -(d-1)/2\le a\le (d-1)/2$. Then $a=-u$, and so we decode into $\sigma +a=\pi_i$.

\subsection{First recursive construction}

Let $C$ be an $(n, d)$ PA of size $M$, and let $r\ge 2$ be an integer.  We
define an $(rn, rd)$ PA, $C_r$, of size $M^r$ as follows:
for each multi-set of $r$ code words from $C$,
\[(\pi_1^{(j)}, \dots, \pi_n^{(j)}), j=0, 1,\dots , r-1,\]
let
\[\rho_j=(r\pi_1^{(j)}-j, \dots, r\pi_n^{(j)}-j), j=0, 1,\dots , r-1,\]
and include $(\rho_0|\rho_1|\dots |\rho_{r-1})$ as a codeword in $C_r$.
It is clear that under this construction the distance between any two distinct $\rho_j$,
$\rho_{j'}$ is at least $rd$.
It is also easy to check that $(\rho_0|\rho_1|\dots |\rho_{r-1})\in S_{rn}$.
Hence $|C_r|=P(n,d)^r.$ In particular, we get the following bound.

\begin{theorem}
If $n> d$ and $r\ge 2$, then
\[P(rn, rd)\ge P(n,d)^r.\]
\end{theorem}

\subsection{Second recursive construction}

For a permutation $\pi=(\pi_1,\pi_2,\ldots ,\pi_n)\in S_n$ and an integer $m$, $1\le m \le n+1$
define
\[\varphi_m(\pi)=(m,\pi_1',\pi_2',\ldots,\pi_n')\in S_{n+1}\]
by
\[\begin{array}{ll}
\pi_i'=\pi_i & \mbox{if }\pi_i\le m, \\
\pi_i'=\pi_i+1 & \mbox{if }\pi_i> m.
\end{array}\]

 Let $C$ be an $(n,d)$ PA,
and let
\[1\le s_1<s_2<\cdots <s_t\le n+1\]
 be integers.
Define
\[C[s_1,s_2,\ldots ,s_t]=\{\varphi_{s_j}(\pi) \mid 1\le j \le t,\ \pi\in C\}.\]

\begin{theorem}
\label{tc21}
If $C$ is an $(n,d)$ PA of size $M$ and
\[s_j+d\le s_{j+1}\mbox{ for }1\le j \le t-1,\]
then $C[s_1,s_2,\ldots ,s_t]$ is an $(n+1,d)$ PA of size $tM$.
\end{theorem}

\begin{theorem}
\label{tc22}
If $C$ is an $(n,d)$ PA of size $M$ and $n\le 2d$, then
$C[d]$ is an $(n+1,d+1)$ PA of size $M$.
\end{theorem}

Proof. 
If $j>j'$, then
\[d_{\max}(\varphi_{s_j}(\pi),\varphi_{s_{j'}}(\sigma))\ge s_j-s_{j'}\ge d.\]
Next, consider $j'=j$.
If $\pi,\sigma\in C$, $\pi\ne \sigma$, then w.l.o.g, there exist an $i$ such that $\pi_i\ge \sigma_i+d$. Hence
\[d_{\max}(\varphi_{s_j}(\pi),\varphi_{s_j}(\sigma)) \ge \left\{
\begin{array}{ll}
\pi_i-\sigma_i+1> d &\mbox{if } \pi_i> s_j\ge \sigma_i,\\
\pi_i-\sigma_i\ge d &\mbox{otherwise.}
\end{array}
\right. \]
This proves Theorem \ref{tc21}.
To complete the proof of Theorem \ref{tc22} we note that
\[\pi_i\ge \sigma_i+d\ge d+1>d,\]
and
\[\sigma_i\le \pi_i-d\le n-d\le d.\]
Hence $\pi_i> d\ge \sigma_i$ and so
\[d_{\max}(\varphi_{s_j}(\pi),\varphi_{s_j}(\sigma))\ge d+1.\]
 \medskip

The constructions imply bounds on $P(n,d)$. First, choosing $t=\lfloor n/d\rfloor +1$,
$s_t=n+1$ and $s_j=(j-1)\lfloor n/d\rfloor+1$ for $1\le j \le t-1$, we get the following bound.

\begin{theorem}
\label{c2b1}
If $n>d\ge 1$, then
\[P(n+1,d)\ge \left( \left\lfloor \frac{n}{d}\right\rfloor+1\right)P(n,d).\]
\end{theorem}

\begin{example}
\label{recb}
In Example \ref{expb} we showed that the explicit construction implied that $P(2a,2)\ge (a!)^2$.
Combining Theorem \ref{c2b1} and search, we can improve this bound. We have found that $P(7,2)\ge 582$, see the
table at the end of the next section. From repeated use of Theorem \ref{c2b1} we get
\[P(2a,2)\ge (a(a-1)\cdots 5)^2\cdot 4\,  P(7,2)\ge \frac{97}{24}(a!)^2.\]
\end{example}

Theorem \ref{tc22} implies the following bound.

\begin{theorem}
\label{tr1}
If $d<n\le 2d$, then
\[P(n+1,d+1)\ge P(n,d).\]
\end{theorem}

Theorem \ref{tr1} shows in particular that for a fixed $r$, 
\begin{equation}
\label{ba}
P(d+1+r,d+1)\ge P(d+r,d)\mbox{ for }d\ge r.
\end{equation} 
We will show that $P(d+r,d)$ is bounded. We show the following theorem.

\begin{theorem}
\label{con2}
For fixed $r$, there exist constants $c_r$ and $d_r$ such that $P(d+r,d)=c_r$ for $d\ge d_r$. Moreover,
\begin{equation}
\label{bc}
c_r\le  2^{2r}\, (2r)!
\end{equation}
and 
\begin{equation}
\label{bd}
d_r\le 1+(2r-1)c_r-r.
\end{equation} 
\end{theorem}

Remark. The main point of Theorem \ref{con2} is the existence of $c_r$ and $d_r$. The actual bounds given
are probably quite weak in general. For example, Theorem \ref{con2} gives the bounds $c_1\le 8$ and $d_1\le 8$.
In Theorem \ref{p1} below, we will show that $c_1=3$ and $d_1=2$. Theorem \ref{con2} gives $c_2\le 384$ and $d_2\le 1151$,
whereas numerical computation indicate that $c_2=9$ and $d_2=5$.

We split the proof of Theorem \ref{con2} into three lemma.

\begin{lemma}
\label{lem1}
If $d\ge r$, then $P(d+r,d)\le 2^{2r}\, (2r)!$.
\end{lemma}
Proof. 
Suppose that there exists an $(d+r,d)$ PA $C$ of size $M>2^{2r}\, (2r)!$.
We call the integers
\[1,2,\ldots,r\mbox{ and }d+1,d+2,\ldots ,d+r\]
 \emph{potent}, the first $r$ \emph{smaller potent}, the last $r$ \emph{larger potent}.
Two potent integers are called {\em equipotent} if both are smaller potent or both are larger potent.
If the distance between two permutations $(\pi_1,\pi_2,\ldots , \pi_n)$, $(\rho_1,\rho_2,\ldots ,\rho_n)$ is at least $d$, then
there exists some position $i$ such that, w.l.o.g, $\pi_i-\rho_i\ge d$, Then $\pi_i$ is a larger potent element and $\rho_i$ is smaller potent.
Each permutation in $S_{d+r}$ contains $2r$ potent elements and we call the set of positions of these the \emph{potency support}
$\chi(\pi)$ of the permutation, that is, the potency support of $\pi$ is
\[\chi(\pi)=\{i\mid 1\le \pi_i \le r\}\cup \{i \mid d+1\le \pi_i\le d+r\}.\]
The potency support of $C$ is the union of the potency support of the permutations in $C$, that is
\begin{align*}
\chi(C)=&\{i\mid 1\le \pi_i \le r\mbox{ for some }\pi\in C\} \\
    & \cup \{i \mid d+1\le \pi_i\le d+r\mbox{ for some }\pi\in C\}.
\end{align*}

Let $\pi\in C$. For each $\rho\in C$, $\rho\ne \pi$, we have $d(\pi,\rho)\ge d$. Hence there exists some $i\in \chi(\pi)$
such that $\rho_i$ is potent. Therefore, the set
\[\{(\rho,i) \mid \rho\in C \mbox{ and }i\in \chi(\pi)\}\]
contains at least $2r+(M-1)>M$ elements. Hence there is an $i\in \chi(\pi)$ such that
\[|\{\rho\in C \mid \rho_i\mbox{ is potent}\}|> M/(2r)> 2^{2r} (2r-1)!.\]
Since
\begin{eqnarray*}
 \{\rho\in C \mid \rho_i\mbox{ is potent}\} &=& \{\rho\in C \mid \rho_i\mbox{ is smaller potent}\}\\
 && \cup \{\rho\in C \mid \rho_i\mbox{ is larger potent}\},
\end{eqnarray*}
there exists a subset $C_1\subset C$ such that
\[|C_1|>2^{2r-1} (2r-1)!\]
 and the elements in position $i_1=i$ are equipotent.

We can now repeat the procedure. Let $\pi\in C_1$. There must exist an $i_2\in \chi(\pi)\setminus \{i_1\}$ such that
\[|\{\rho\in C_1 \mid \rho_{i_2}\mbox{ is potent}\}|\ge |C_1|/(2r-1)> 2^{2r-1} (2r-2)!.\]
Hence we get subset $C_2\subset C_1$ such that
\[|C_2|>2^{2r-2} (2r-2)!\]
 and the elements in position $i_2$ are equipotent (and the elements
in position $i_1$ are equipotent).

Repeated use of the same argument will produce for each $j$, $1\le j \le 2r$ a set $C_j$ such that
\[|C_j|> 2^{2r-j}\, (2r-j)!\]
and for $j$ positions $i_1,i_2,\ldots i_j$, the elements in those positions are all equipotent.
In particular, $|C_{2r}|>1$, all permutations in $C_{2r}$ have the
same potency support $\{i_1,i_2,\ldots ,i_{2r}\}$, and for each of these positions, all the elements in that position are equipotent.
This is a contradiction since the distance between two such permutations must be less than $d$.
Hence the assumption that a PA of size larger than $2^{2r}\, (2r)!$ exists leads to a contradiction.
 \medskip

Lemma \ref{lem1} combined with (\ref{ba}) proves the existence of $c_r$ and $d_r$ and gives the bound (\ref{bc}).

\begin{lemma}
\label{s1}
If $C$ is a $(d+r,d)$ PA of size $M$ where
\[d>r\mbox{ and }d+r>|\chi(C)|,\]
 then there exists a $(d-1+r,d-1)$ PA of size $M$.
In particular, if $M=P(d+r,d)$, then
\[P(d-1+r,d-1)=P(d+r,d).\]
\end{lemma}
Proof. 
Replace all elements in range $r+1,r+2,\ldots ,d$ in the permutations of $C$ by a star $*$ which will denote "unspecified".
The permutations in $C$ is transformed into \emph{vectors} containing the potent elements and $d-r$ stars.
Note that if we replace the unspecified elements in each vector by the integers $r+1,r+2,\ldots ,d$ in some order, we get a permutation,
and the distance between two such permutations will be at least $d$ since we have not changed the potent elements.

Since the length $d+r$ of $C$ is larger than $|\chi(C)|$, there exists a position where all the vectors contains a star.
Remove this position from each vector and reduce all the larger potent elements by one. This given a set of $M$ vectors of length $d-1+r$
and such that the distance between any two is at least $d-1$. Replacing the $d-1-r$ stars in each vector by $r+1,r+1,\ldots, d-1$ in some order,
we get an $(d-1+r,d-1)$ PA of size $M$.

If $M=P(d+r,d)$, then we get
\[P(d-1+r,d-1)\ge P(d+r,d).\]
Since $P(d-1+r,d-1)\le P(d+r,d)$ by (\ref{ba}), the lemma follows.
\medskip

\begin{lemma}
\label{s2}
If $C$ is a $(d+r,d)$ PA of size $M$ and $d\ge r$, then
\[|\chi(C)|\le M(2r-1)+1.\]
\end{lemma}
Proof. 
Each permutation has potency support of size $2r$.
The potency support of any two permutations in $C$ must overlap since their distance is at least $d$.
Hence each permutation after the first will contribute at most $2r-1$ new elements to the total potency support.
Therefore,
\[|\chi(C)|\le 2r +(M-1)(2r-1).\]
\medskip

Remark. By a more involved analysis, we can improve this bound somewhat. For example, we see that two new permutations
can contribute at most $4r-3$ to the total support.

We can now complete the proof of Theorem \ref{con2}. Let $C$ be a $(d+r,r)$ code of size $c_r$. By Lemma \ref{s2}, 
$|\chi(C)|\le c_r(2r-1)+1$. If $d> 1+c_r(2r-1)-r$, then $d+r>|\chi(C)|$. Hence, by Lemma \ref{s1}, $P(d-1+r,d-1)=P(d+r,d)$.
Therefore, $d_r\le 1+c_r(2r-1)-r$, that is, (\ref{bd}) is satisfied. This completes the proof of Theorem \ref{con2}.

\begin{theorem}
\label{p1}
We have $P(d+1,d)=3$ for $d\ge 2$.
\end{theorem}
Proof. 
We use the same notation as in the proof of Lemma \ref{s1}.
Let $C$ be an $(d+1,d)$ PA. The only potent elements are $1$ and $n$.
W.lo.g. we may assume the first permutation in $C$ is $(1,n,*,*,\ldots)$ where $*$ denotes some unspecified integer in the range $2,3,\ldots ,d$. W.l.o.g,
a second permutation has one of three forms:
\[(n,1,*,*,\ldots), \, (n,*,1,*,\ldots), \, (*,1,n,*,\ldots).\]
We see that if the second permutation is of the first form, there cannot be more permutations. If the second permutation is of the
form $(n,*,1,*,\ldots)$,
then there is only one possible form for a third permutation, namely $(1,*,n,*,\ldots)$. Hence we see that $P(d+1,d)\le 3$ and
that $P(d+1,d)=3$ for $d\ge 2$.
 \medskip

To determine $P(d+r,d)$ along the same lines for $r\ge 2$ seems to be difficult because of the many cases that have to be considered.
Even to determine $P(d+2,d)$ will involve a large number of cases. For example for the second permutation there are
138 essentially different possibilities for the four positions in the potency support of the first permutation. For each of these there are
many possible third permutations, etc.

\subsection{Encoding/decoding of some PA constructed by the second recursive construction}
Suppose we start with the PA
\[C_{d}=\{(1,2,3,\ldots,d)\}.\]
For $\nu=d,d+1,\ldots,n-1$ let
\[C_{\nu+1}=C_{\nu}[1,\nu+1].\]
 Then $C_n$ is an $(n,d)$ PA of size $2^{n-d}$.
For some applications, we may want to map a set of binary vectors to a permutation array.
One algorithm for mapping a binary vector $(x_1,x_2,\ldots ,x_{n-d})$ into $C_n$ would be to use the recursive construction of $C_n$
by mapping $(x_1,x_2,\ldots,x_i)$ into a permutation $\pi$ in $C_{d+i}$. Recursively, we can then map
$(x_1,x_2,\ldots,x_i,0)$ to $f_1(\pi)$ and $(x_1,x_2,\ldots,x_i,1)$ to $f_{d+i+1}(\pi)$.

However, there is an alternative algorithm which requires less work.
Retracing the steps of the construction, we see that given some initial part of length less than $n-d$ of a permutation in $C_{n}$, there
are exactly two possibilities for the next element, one "larger" and one "smaller". More precisely, induction shows that if the initial part
of length $i-1$ contains exactly $t$ "smaller" elements, then element number $i$ is either $t+1$ (the "smaller") or $n-i+t+1$ (the "larger").
This is the basis for a simple mapping from $Z_2^{n-d}$ to $C_{n}$. We give this algorithm in Figure \ref{Gn}.

\begin{figure}[htbp]
\begin{tabbing}
xx\=xx\=xx\= \kill \\
{\bf Input:}  $(x_1,\dots,x_{n-d}) \in Z_2^{n-d}$\\
{\bf Output:} $(\pi_1,\dots,\pi_{n}) \in C_{n}$\\
\>{\bf for} $i\leftarrow n-d+1$ {\bf to} $n$ {\bf do} $x_i\leftarrow 0;$\\
\>$t \leftarrow  0$;  //* $t$ is the number of zeros seen so far.*//\\
\>{\bf for} $i \leftarrow 1$ {\bf to} $n$ {\bf do}\\
\>\> {\bf if} $x_{i}=0$\\
\>\>\>{\bf then }\{$\pi_i \leftarrow t+1$; $t \leftarrow t+ 1$;\}\\
\>\>\>{\bf else} \{$\pi_i \leftarrow n-i+t + 1$;\}\\
\end{tabbing}
\caption{Algorithm mapping $Z_2^{n-d}$ to $C_{n}$} \label{Gn}
\end{figure}

We see that the difference between the larger and the smaller element in position $i\le n-d$ is $n-i$.
Hence we can recover from any error of size less than $(n-i)/2$ by choosing the closest of the two possible values,
and the corresponding binary value. We give the decoding algorithm in Figure \ref{Gd}.

\begin{figure}[htbp]
\begin{tabbing}
{\bf Input:} $(\pi_1,\dots,\pi_{n}) \in [n]^n$\\
{\bf Output:} $ (x_1,\dots,x_{n-d}) $\\
\hspace{.2cm}  $t\leftarrow 0$;  //* $t$ is number of zeros determined. *//\\
\hspace{.2cm} {\bf for} $i\leftarrow1$ {\bf to} $n-d$ {\bf do}\\
\hspace*{.6cm}\,\, {\bf if} \= $\pi_i< (n-i)/2+t+1$  \\
\>{\bf then }\=\{$x_i \leftarrow 0$; $t\leftarrow t+1$;\}\\
\>{\bf else} \=\{$x_i \leftarrow 1$;\}\\

\end{tabbing}
\caption{Decoding algorithm recovering the binary preimage from a corrupted permutation in $C_{n}$. } \label{Gd}
\end{figure}

Without going into all details, we see that we can get a similar mapping from $q$-ary vectors. Now we start with the PA
\[C_{(q-1)d}=\{(1,2,3,\ldots,(q-1)d)\}.\]
 For $(q-1)d\le \nu\le n-1$ let $s_j=(j-1)\lfloor \nu/(q-1)\rfloor+1$ for
$1\le j \le q-1$ and $s_q=\nu+1$. Let
\[C_{\nu+1}=C_{\nu}[s_1,s_2,\ldots ,s_q].\]
 Then $C_n$ is an $(n,d)$ PA of size $q^{n-(q-1)d}$. Encoding and decoding correcting errors of size
at most $(d-1)/2$, based on the recursion, is again relatively simple.

\section{Further bounds on $P(n, d)$}
\label{sec-bounds}

\subsection{General bounds}

Since $d_{\max}(\pi,\sigma)\le n-1$ for any two distinct permutations in $S_n$, we have $P(n,n)=1$. Therefore, we only consider $d<n$.

Since the spheres of radius $d$ in $S_n$ all have size $V(n,d)$, we can get a Gilbert type lower bound on $P(n,d)$.

\begin{theorem}
\label{Gbound}
For $n>d\ge 2$ we have
\[P(n,d) \ge \frac{n!}{V(n,d-1)}.\]
\end{theorem}
Proof. 
It is clear that the following greedy algorithm produces a permutation array with cardinality at least $n!/ V(n,d-1)$.
\begin{enumerate}
\item Start with any permutation in $S_n$.
\item  Choose a permutation whose distance is at least $d$ to all previous chosen permutations.
\item Repeat step 2 as long as such a permutation exists.
\end{enumerate}
Let $C$ be the permutation array produced by the above greedy algorithm. Once the algorithm stops, $S_n$ will
be covered by the $|C|$ spheres of radius $d-1$ centered at the code words in $C$. Thus $n! \le |P| \cdot  V(n,d-1)$ which implies our claim.
\medskip

Similarly, since the spheres $V(n,\lfloor (d-1)/2\rfloor)$ are disjoint, we get the following Hamming type upper bound.
\begin{theorem}
\label{Hbound}
If $n> d \ge 1$, then
\[P(n,d) \le \frac{n!}{V(n,\lfloor (d-1)/2\rfloor)}.\]
\end{theorem}

If $n\le 2d$ and $d$ is even, we can combine the bound in Theorem \ref{Hbound} with Theorem \ref{tr1} to get the following bound
which is stronger than the ordinary Hamming bound, at least in the cases we have tested.
\begin{theorem}
\label{Hbound2}
If $d$ is even and $2d\ge n > d \ge 2$, then
\[P(n,d) \le \frac{(n+1)!}{V(n+1,d/2)}.\]
\end{theorem}

\begin{example}
For $n=11$ and $d=6$, Theorem \ref{Hbound} gives
\[P(11,6)\le \left\lfloor \frac{11!}{V(11,2)}\right\rfloor =\left\lfloor \frac{11!}{11854}\right\rfloor = 3367\]
whereas Theorem \ref{Hbound2} gives
\[P(11,6)\le \left\lfloor \frac{12!}{V(12,3)}\right\rfloor =\left\lfloor \frac{12!}{563172}\right\rfloor = 850.\]
\end{example}

Remark. We can of course use Theorem \ref{tr1} repeatedly $r$ times and then Theorem \ref{Hbound} to get
\[P(n,d) \le \frac{(n+r)!}{V(n+r,\lfloor (d+r-1)/2\rfloor)}\]
for all $r\ge 0$. However, it appears we get the best bounds for $r=1$ when $d$ is even and $r=0$ when $d$ is odd.\medskip

In general, no simple expression of $V(n,d)$ is known. A survey of known results as well as a number of new results on $V(n,d)$
were given by Kl{\o}ve \cite{Klove08}. Here we briefly give some main results.

As observed by Lehmer \cite{L70},
$V(n,d)$ can be expressed as a permanent.
The permanent of an $n \times n$ matrix $A$  is defined by
\[{\rm per} A = \sum_{\pi \in S_n}  a_{1,\pi_1} \cdots a_{n,\pi_n}.\]
In particular, if $A$ is a $(0,1)$-matrix, then
\[{\rm per}A = |\{\pi\in S_n : a_{i, \pi_i}=1 \mbox{ for all }i\}|.\]

Let $A^{(n,d)}$ be the $ n \times n$ matrix with $a^{(n,d)}_{i,j}=1$ if $|i-j|\le d$ and $a^{(n,d)}_{i,j}=0$ otherwise.

\begin{lemma} $V(n,d)={\rm per} A^{(n,d)}$.
\end{lemma}
 Proof. 
\begin{eqnarray*}
V(n,d) &=&  |\{\pi \in S_n : d_{\max}(\iota,\pi) \le d\}| \\
  &=& |\{\pi \in S_n : |i-\pi_i| \le d \mbox{ for all }i\}| \\
  &=& |\{\pi \in S_n : a^{(n,d)}_{i,\pi_i}=1 \mbox{ for all }i\}| \\
 &=& {\rm per} A^{(n,d)}.
\end{eqnarray*}

For fixed $d$, $V(n,d)$ satisfies a linear recurrence in $n$. A proof is given in \cite{Stanley} (Proposition 4.7.8 on page 246).
For $1\le d\le 3$ these recurrences were determined explicitly by Lehmer \cite{L70}, and for $4\le d \le 6$ by Kl{\o}ve \cite{Klove08}.
In particular, this implies that
\[\lim_{n\rightarrow\infty}V(n,d)^{1/n}=\mu_d,\]
where $\mu_d$ is the largest root of the minimal polynomial corresponding to the linear recurrence of $V(n,d)$.
Lehmer \cite{L70} determined $\mu_d$ approximately for $d=1,2,3$ and Kl{\o}ve \cite{Klove08} for $d\le 8$.

For an $n\times n$ $(0,1)$-matrix it is known (see Theorem 11.5 in \cite{LW03}) that
\[ {\rm per}A\le \prod_{i=1}^n (r_i!)^{1/r_i},\]
where $r_i$ is the number of ones in row $i$.

For $A^{(n,d)}$ we clearly have $r_i\le 2d+1$ for all $i$. Hence
\begin{equation}
\label{Vupper}
 V(n,d)\le [(2d+1)!]^{n/(2d+1)}\mbox{ for all }n
\end{equation}
and
\[ \mu_d\le [(2d+1)!]^{1/(2d+1)}.\]
In Table \ref{tab3} we give $\mu_d$ and this upper bound.

\begin{table}[htb]
\caption{$\mu_d$ and its upper bound. \label{tab3}}
\[\begin{array}{|c|c|c|c|c|}\hline
d & \mu_d & [(2d+1)!]^{1/(2d+1)} & \mu_d/(2d+1) \\ \hline
1 & 1.61803 & 1.81712 & 0.53934 \\
2 & 2.33355 & 2.60517 & 0.46671 \\
3 & 3.06177 & 3.38002 & 0.43739 \\
4 & 3.79352 & 4.14717 & 0.42150 \\
5 & 4.52677 & 4.90924 & 0.41152 \\
6 & 5.26082 & 5.66769 & 0.40468 \\
7 & 5.99534 & 6.42342 & 0.39969 \\
8 & 6.73016 & 7.17704 & 0.39589 \\ \hline
\end{array}\]
\end{table}

We note that for large $d$, $\mu_d/(2d+1)\approx 1/e$.

Combining Theorem \ref{Gbound} and (\ref{Vupper}) we get

\begin{corollary}
For $n>d\ge 1$, we have
\[P(n,d) \geq \frac{n!}{[(2d-1)!]^{n/(2d-1)} }.\]
\end{corollary}

\subsection{Table of bounds on $P(n,d)$ }

We have used the following greedy algorithm to find an $(n,d)$ PA $C$: Let the identity permutation in $S_n$ be the first permutation in $C$.
For any set of permutations chosen,
choose as the next permutation in $C$ the lexicographically next permutation in $S_n$ with distance at least $d$ to the chosen permutations
in $C$
if such a permutation exists.
The size of the resulting PA is of course a lower bound on $P(n,d)$.

The lower bounds in Table \ref{tab4} were in most cases found by this greedy algorithm.
For $n=8$, $d=5$, the greedy algorithm gave a PA of size 26.
However,
\[P(8,5) \ge P(7,4) \ge 28\]
by Theorem \ref{tr1}. Similarly,
\[P(10,7)\ge P(9,6)\ge P(8,5) \ge 28.\]
Some other of the lower bounds are also determined using Theorem \ref{tr1}. They are marked by $*$.
The upper bound is the Hamming type bound in Theorem \ref{Hbound} or it's modified bound in Theorem \ref{Hbound2}.
Since $P(n,1)=n!$ for all $n$, this is not included in the table.

\begin{table}[htb]
\caption{Bounds on $P(n,d)$. \label{tab4}}
\[\begin{array}{l|ccc}
    & d=2    & d=3   & d=4    \\ \hline
n=d+1 &  3    &  3    &  3    \\
n=d+2 & 6-24    &  9   & 9-12    \\
n=d+3 & 29-120  & 20-34  & 28-43  \\
n=d+4 & 90-720  & 84-148 & 68-166  \\
n=d+5 & 582-5040 &401-733 & 283-4077 \\ \hline
\end{array}\]
\[\begin{array}{l|ccc}
    & d=5    & d=6   & d=7  \\ \hline
n=d+1 &  3    & 3    &  3  \\
n=d+2 & 9-12   & 9-18  & 9-18  \\
n=d+3 & 28^*-43 & 28^*-60  & 28^*-60 \\
n=d+4 & 95-166  & 95^*-216 & 95^*-216  \\
n=d+5 & 236-714 & 236^*-850 & 236^*-850  \\ \hline
\end{array}\]
\end{table}

\section{Conclusion}\label{conclusion:sec}

We give a number of constructions of permutations arrays under the Chebyshev distance, some with efficient decoding algorithms.
We also consider an explicit mapping of vectors to permutations with efficient encoding/decoding.
Finally, we give some bounds on the size of PAs under the Chebyshev distance.

\end{document}